\documentclass{ifacconf}

\usepackage{graphicx}      
\usepackage{natbib}        
\usepackage{mathtools}
\usepackage{amsmath}
\usepackage{breqn}
\usepackage{tabularx}
\usepackage{tikz}
\usetikzlibrary{shapes,arrows,positioning,calc}
\begin{document}

\tikzset{
block/.style = {draw, fill=white, rectangle, minimum height=2.5em, minimum width=2.5em},
block2/.style = {draw, fill=white, rectangle, minimum height=2cm, minimum width=1cm},
lblock/.style = {draw, fill=white, rectangle, minimum height=8em, minimum width=3em},
tmp/.style  = {coordinate}, 
point/.style  = {coordinate},
sum/.style= {draw, fill=white, circle, node distance=1cm},
input/.style = {coordinate},
output/.style= {coordinate},
pinstyle/.style = {pin edge={to-,thin,black}
}
}

\begin{frontmatter}

\title{Modelling of Variable Speed Hydropower\\for Grid Integration Studies\thanksref{footnoteinfo}} 

\thanks[footnoteinfo]{This work was supported by the Research council of Norway under Grant 257588 and by the Norwegian  Research  Centre for Hydropower Technology (HydroCen).}

\author[First]{T.I. Reigstad} 
\author[Second]{K. Uhlen} 

\address[First]{Norwegian University of Science and Technology (NTNU), NO-7491 Trondheim, Norway (email: tor.inge.reigstad@ntnu.no)}
\address[Second]{Norwegian University of Science and Technology (NTNU), NO-7491 Trondheim, Norway (email: kjetil.uhlen@ntnu.no)}

\begin{abstract}                
This paper proposes a hydraulic model based on the Euler turbine equations suitable for the purpose of grid integration studies of variable speed hydropower (VSHP). The work was motivated by the need to assess how the dynamic performance might change when a hydropower plant is operated at variable speed. The \textit{Euler} model considers the water flow dependency on the turbine rotational speed and calculates the turbine power as a non-linear function of water flow, turbine rotational speed and guide vane opening. A waterway model is included, based on the 1-D momentum and continuity balance for a water-filled elementary pipe to simulate water hammer, mass oscillation and tunnel losses. These detailed and accurate models are necessary for recognising possible limitations in the hydraulic system, to model the turbine power and rotational speed correctly and thereby to be able to maximise power delivery for system control purposes. All \textit{Euler} model parameters can be derived from the physical dimensions of the turbine and waterway, ensuring easy implementation. State-space representation of the \textit{Euler} model is approximated by utilising a lumped-parameter equivalent of the penstock dynamics. Dynamic simulations and eigenvalue analysis  show the strength of the \textit{Euler} model compared to conventional hydropower models. 
\end{abstract}


\end{frontmatter}

\section{Introduction}
\begin{figure}[!t]
\centering
    \includegraphics[scale=0.8]{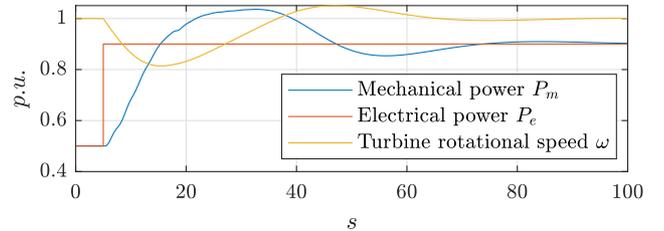}
    \caption{Dynamic response when the VSHP output power is increased from 0.5~p.u. to 0.9~p.u. at $t=5s$}\label{fig:p}
\end{figure}

Variable speed hydropower (VSHP) is a suitable source for delivering additional ancillary services to the grid by actively utilising the stored kinetic energy in the turbine and generator. By allowing the turbine rotational speed to deviate temporarily from its optimal speed, the VSHP can vary its output power quickly due to the converter technology, see \cite{basic2018high}. In the first few seconds after a step response at the output power reference, the energy is delivered to or from the kinetic energy in the turbine and generator (Figure \ref{fig:p}). Subsequently, the governor will react to the deviation in the turbine rotational speed and adjust the guide vane opening and thereby the mechanical power $P_m$ to regain the optimal speed of the turbine. With that, the VSHP can contribute more effectively to primary frequency control and the maintaining of grid stability. The VSHP plants will be able to provide fast frequency reserves in both production and in pumping mode, and the efficiency and operating range will potentially be wider than for conventional hydropower and other variable renewable sources without storage, see \cite{valavi2018variable}.


The turbine and hydraulic system, including waterways, will experience new and different operating conditions when running at variable speed. When analysing the power system impacts and potential benefits of variable speed operation, we need to know which dynamic constraints and limitations concerning the technology need to be taken into account. Thus, in this context, there are several good reasons for revisiting the modelling and analysis of hydropower plants, which is the aim of this paper. A sufficiently detailed model of the system is needed to investigate the interactions between the VSHP plant and the grid: How can variable speed operation benefit the security and flexibility in power system operation? How can we explore the control possibilities from a system perspective while considering the limitations given by the water/turbine system? This requires the development of non-linear time-domain simulation models that include tunnels, with physical constraints on water flows, turbine, governor, generator with the magnetising system, generator-side converter and grid-side converter, and a representative test grid. This paper will focus on the turbine side of the generator by comparing four different turbine models, two of them including a waterway model, to examine how accurate the models are when subjected to large variations in turbine rotational speed.

The hydraulic system is modelled in different ways, depending on application and performance requirements. For example, the conduits in the hydraulic system can be modelled as electrical equivalent circuits, as presented in \cite{souza1999study, nicolet2007hydroacoustic} or as travelling waves, as suggested in \cite{demello1992hydraulic}. These models can consider both penstock, surge tank and tunnel dynamics and losses. Hydroelectric turbine-governor simulation models for commercial simulation programs are reviewed in \cite{koritarov2013review}. Except for the Hygovm model, most of the existing models only consider the water starting time when modelling the conducts. 

The simplest turbine models assume that mechanical power equals the water flow in per unit, see \cite{koritarov2013review} and \cite{kundur1994power}. Hygov and similar models in \cite{koritarov2013review} do also consider variation in the head, the damping and the losses by subtracting the no-load water flow. These models are linearised at the operational point, as seen in \cite{sarasua2015dynamic} or a linearized model can be found from the turbine characteristic chart at the operational point, see \cite{fang2008basic}. Hill diagrams are used to include the turbine efficiency in \cite{belhadji2011modeling}. Turbine characteristics are used in \cite{pannatier2010investigation} and \cite{padoan2010dynamical} to find the torque and flow when rotational speed and guide vane opening are given. The Euler turbine equations of \cite{nielsen2015simulation} also consider the rotational speed. A one-dimensional numerical model of a Francis turbine based on the Euler equations, which includes a waterway model is presented in \cite{giosio2016physics}. This model is tuned with test data and utilises look-up tables to find churning losses and is utilised in a VSHP model with a doubly-fed induction machine (DFIM) in \cite{nag2018dfim}.  


\cite{fang2008basic} present basic mathematical models for typical hydroelectric power plants and discuss how a turbine speed governor should be optimally tuned and how the physical dimensions of the waterway affect the dynamics. In \cite{mohanpurkar2018real}, a transient stability analysis including VSHP with an elastic water column model, turbine model and DFIM is performed. 

This paper is structured as follows: The models assessed for application to the analysis of variable speed hydropower models are described in Section \ref{VarialbeSpeedModels}. Section \ref{ComparisonModels} presents the simulation result for four different hydraulic models and discusses the differences and strength of the models. The conclusions are summed up in Section \ref{Conclusion}.

\section{Variable Speed Hydropower Models}\label{VarialbeSpeedModels}

\subsection{Waterway Model}

To investigate how the hydraulic system and electrical system affect each other, the waterway model has to consider both water hammering and friction losses. Inelastic water column models are adequate for short penstocks; however, with longer penstocks, the effects of travelling waves and thereby the elasticity of the steel and the compressibility of the water must be considered. The derivation of the model starts with the 1-D momentum and continuity balance for a water-filled elementary pipe of length $dx$ given as a partial differential equation in \cite{wylie1993fluid} which can be represented as \eqref{hypPar}, see \cite{pico2012analysis}.

\begin{equation}\label{hypPar}
    \begin{split}
     \frac{\delta H}{\delta x} + \frac{1}{g_r A} \cdot \frac{\delta Q}{\delta t} + \frac{f}{2g_rDA^2} \cdot Q|Q| &= 0\\
     \frac{\delta Q}{\delta x} + \frac{g_rA}{a^2} \cdot \frac{\delta H}{\delta t} &= 0  
    \end{split}
\end{equation}

The cross-section is $A$, $Q$ is the discharge, $H$ is the head, $g_r$ is the gravity, $D$ is the pipe diameter, $f$ is the local loss coefficient, and $a$ is the wave speed. Equation \eqref{hypPar} can be per unitised to \eqref{hypParpu} by using the per unitised definitions of head and flow given in \eqref{hqpu}, see \cite{pico2012analysis}.

\begin{equation}\label{hqpu}
    \begin{split}
     h_{p.u.}=\frac{H}{H_R}, \quad     q_{p.u.}=\frac{Q}{Q_R}
    \end{split}
\end{equation}

\begin{equation}\label{hypParpu}
    \begin{split}
     \frac{\delta q}{\delta x} + \frac{g_r AH_R}{Q_R a^2} \cdot \frac{\delta h}{\delta t} &= 0  \\
     \frac{\delta h}{\delta x} + \frac{Q_R}{H_R g_r A} \cdot \frac{\delta q}{\delta t} + \frac{fQ_R^2}{2g_r H_R DA^2} \cdot q|q| &= 0
    \end{split}
\end{equation}

By using Laplace transformation and neglecting the friction losses, \eqref{hypParpu} can be solved to find the transfer function of the flow rate $q_U$ and the water pressure $h_U$ of the upstream inlet as a function of the downstream outlet flow rate $q_D$ and the water pressure $h_D$ \cite{nicolet2007hydroacoustic}:

\begin{multline}\label{hypParMa}
    \begin{bmatrix}
        h_U \left(s\right) \\
        q_U \left(s\right)
    \end{bmatrix}
    =\\
    \begin{bmatrix}
        \cosh{\left(zT_e\right)} & -Z_c \sinh{\left(zT_e\right)}\\
        -\frac{1}{Z_c} \sinh{\left(zT_e\right)} & \cosh{\left(zT_e\right)} 
    \end{bmatrix}
    \begin{bmatrix}
        h_D \left(s\right) \\
        q_D \left(s\right)
    \end{bmatrix}
\end{multline}

where the characteristic impedance is given as

\begin{equation}\label{tw}
    Z_c = \frac{T_w}{T_e} \frac{z}{s}, \quad \text{where} \quad T_w = \frac{L}{g_r A} \frac{Q_R}{H_R}, \quad T_e = \frac{L}{a}
\end{equation}

The water inertia time constant, also known as the water starting time, is given as $T_w$, $T_e$ is the wave travel time and is the length of the waterway. 

The classic wave solution given below considers both the elastic water hammer theory and the hydraulic losses as a hyperbolic tangent function. The derivation is shown in \cite{brekke1984stability}.

\begin{multline}\label{hoverq}
    \frac{h(s)}{q(s)}= -\frac{T_w}{T_e} \left( 1+\frac{fQ_R}{2DAs}\right)^{1/2}\\ \tanh{\left(\left(s^2+s\frac{fQ_R}{2DA}\right)^{1/2}T_e\right)}
\end{multline}

If the hydraulic friction losses are neglected, \eqref{hoverq} can be simplified to

\begin{equation}\label{hoverqSimpl}
    \frac{h(s)}{q(s)}=-\frac{T_w}{T_e} \tanh{\left(sT_e\right)} = -Z_0 \tanh{\left(sT_e\right)}
\end{equation}

For small variations around the operating point,  $\tanh{\left(sT_e \right)}\approx sT_e$ and \eqref{hoverq} can be simplified to 

\begin{equation}\label{hoverqSimpl2}
    \frac{h(s)}{q(s)}=-T_w s - H_f
\end{equation}

where $H_f$ is the hydraulic friction losses. Neglecting these, \eqref{hoverqSimpl2} becomes

\begin{equation}\label{hoverqSimpl3}
    \frac{h(s)}{q(s)}=-T_w s
\end{equation}

For small-signal stability analysis, \eqref{hoverqSimpl} can be approximated by a lumped-parameter equivalent for $\tanh{\left(sT_e\right)}$. With $n_{max}=0$ and $T_e = 0.5 s$ (inelastic water column), \eqref{tanh} is valid up to approximately 0.1 Hz, for $n_{max}=1$ it is valid up to about 1.0 Hz. \cite{kundur1994power}

\begin{multline}\label{tanh}
    \tanh{\left(sT_e\right)}=\frac{1-e^{-2T_es}}{1+e^{-2T_es}} \\
    \approx \frac{ \displaystyle sT_e \prod_{n=1}^{n_{max}} \left( 1+ \left( \frac{sT_e}{n \pi} \right)^2 \right) }{\displaystyle \prod_{n=1}^{n=\infty} \left( 1+ \left( \frac{2sT_e}{\left(2n-1\right) \pi} \right)^2 \right)}
\end{multline}

The transfer function for surge tanks and air accumulators is given in \cite{xinxin1988hydropower} and rewritten as

\begin{equation}\label{hoverqsurge}
    \frac{h(s)}{q(s)} = \frac{Q_R}{sH_R A_{eqv}} = \frac{1}{sT_s}
\end{equation}

where the surge tank filling time $T_s$ is defined as

\begin{equation}
    T_s = \frac{A_{eqv} H_R}{Q_R}
\end{equation}

The head loss $h_f$ is the pressure drop over a length $l$ of the penstock and can be expressed as \cite{mansoor2000behaviour}

\begin{equation}
    h_f = f_r \frac{l}{d}\frac{v^2}{2g_a}
\end{equation}

where $f_r$ is the friction factor and $d$ is the penstock diameter.

\begin{figure}[!t]
\centering
\begin{tikzpicture}[auto, node distance=1cm,>=latex']
    \node [input, name=qinput] (qinput) {};
    \node [block, right of=qinput, node distance=1.5cm] (Z0) {$Z_0$};
    
    \node [point, right of=Z0, node distance=1.5cm] (Z0r2) {};
    \node [block, below of=Z0r2, node distance=1cm] (gain2) {$2$};
    \node [block, above of=gain2, node distance=2.1cm] (x2){$x^2$};
    \node [block, right of=x2, node distance=1.5cm] (fp1) {$f_{p1}$};
    \node [sum, right of=gain2] (sum1) {};
    \node [block, right of=sum1] (delay) { $e^{-2T_{e1}}$};
    \node [sum, right of=Z0, node distance=4.5cm] (sum2) {};
    \node [sum, right of=sum2,node distance=1cm] (sum3) {};
    \node [output, right of=sum3, node distance=2cm] (output) {};
    \node [point, right of=Z0, node distance=0.7cm] (Z0r) {};
    \node [point, below of=delay, node distance=0.7cm] (delayb) {};
    \node [point, below of=sum2] (delayr) {};
    \node [point, left of=Z0, node distance=1cm] (z) {};
    \node [output, right of=sum3, node distance=1cm] (output) {};
    
    \draw [->] (qinput) -- node[anchor=south,pos=0.0]{$q$} (Z0);
    \draw [->] (z) |- (x2);
    \draw [->] (x2) |- (fp1);
    \draw [->] (Z0)  -- node[name=z3,pos=0.95]{$-$}  (sum2);
    \draw [->]  (Z0r) |- (gain2);
    \draw [->] (sum2) -- (sum3);
    \draw [->] (gain2) -- (sum1);
    \draw [->] (sum1) -- (delay);
    \draw [->] (delay) -| (sum2);
    \draw [->] (fp1)   -| node[pos=0.95] {$-$}(sum3);
    \draw [->]  (delayr) |- (delayb)  -|  node[pos=0.9] {$-$}(sum1);
    
    \draw [-] (sum3) -- node[anchor=south,pos=0.8]  {$h$}(output);
    
    \node [sum, below of=delayr, node distance=1.5cm ] (sum4) {};
    \node [block, left of=sum4] (Cs) {$\frac{1}{sC_s}$};
    \node [sum,below of=z, node distance=2.5cm] (sum5) {};'
    \node [block, below of=Cs, node distance= 1cm] (f0) {$f_{p0}$};
    \node [block, left of=f0,node distance=1.5cm] (mult) {$\times$};
    \node [block, left of=mult,node distance=1.5cm] (abs) {$\mid x \mid$};
 
    \draw [->] (z) -- node[pos=0.95] {$-$} (sum5);
    \draw [->] (sum5) -- (Cs);
    \draw [->] (Cs) -- (sum4);
    \draw [->] (sum4) -| (sum3);
    \draw [->] (f0) -| node[pos=0.9] {$-$} (sum4);
    \draw [->] (abs) -- (mult);
    \draw [->] (mult) -- (f0);
    \node [point, above of=mult,node distance=1cm] (multa) {};
    \draw [->] (multa) -- (mult);
    \node [point, right of=sum5,node distance=0.7cm] (sum5r) {};
    \draw [->] (sum5r) |- (abs);
    
    \node [point, right of=sum4,node distance=0.5cm] (sum4r) {};
    \node [sum, below of=sum4r, node distance=2.2cm ] (sum6) {};
    \node [block, left of=sum6,node distance=1.5cm] (Tw2) {$\frac{1}{sT_{w2}}$};
    \node [block, below of=Tw2, node distance=1cm] (fp2) {$f_{p2}$};
    \node [block, left of=fp2,node distance=1.5cm] (mult2) {$\times$};
    \node [block, left of=mult2,node distance=1.5cm] (abs2) {$\mid x \mid$};
    \node [block, right of=sum6] (cons) {$1$};
    
    \draw [->] (sum6) -- (Tw2);
    \draw [->] (Tw2) -| (sum5);
    
    \draw [->] (sum4r) -- node[pos=0.9] {$-$}(sum6);
    \draw [->] (fp2) -| node[pos=0.9] {$-$} (sum6);
    \draw [->] (abs2) -- (mult2);
    \draw [->] (mult2) -- (fp2);
    \draw [->] (cons) --  (sum6);
    \node [point, above of=mult2,node distance=1cm] (mult2a) {};
    \draw [->] (mult2a) -| (mult2);
    \node [point, left of=Tw2,node distance=4.5cm] (Tw2l) {};
    \draw [->] (Tw2l) |- (abs2);
    
    \draw[red,thick,dotted]  ($(Z0)+(-0.7,-1.8)$)  rectangle ($(sum2)+(0.4,0.5)$) ;
    \draw[blue,thick,dotted] ($(Z0)+(-0.5,-4.0)$)  rectangle ($(sum2)+(0.3,-1.9)$);
    \draw[green,thick,dotted] ($(Z0)+(-0.5,-6.3)$)  rectangle ($(sum2)+(2.1,-4.2)$);
    \draw[orange,thick,dotted] ($(Z0)+(-0.7,0.6)$)  rectangle ($(sum2)+(0.4,2.0)$);
    
    \node [align=left] at ($(Z0)+(2.2,0.2)$) {Penstock dynamics};
    \node [align=center] at ($(Z0)+(1.2,-2.2)$) {Surge tank dynamics};
    \node [align=left] at  ($(Z0)+(1.2,-4.4)$) {Headrace tunnel};
    \node [align=left] at   ($(Z0)+(2.2,1.8)$) {Penstock losses};
    
;
\end{tikzpicture}
\caption{Waterway dynamic model.} \label{figww}
\end{figure}
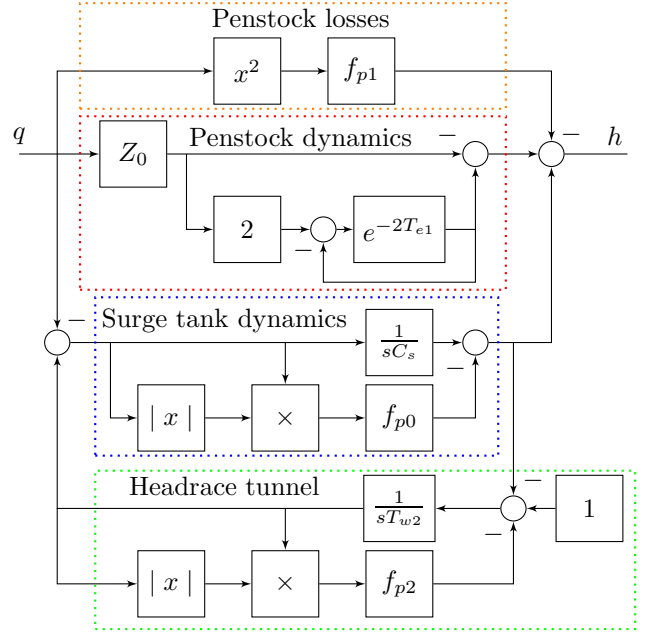

Figure \ref{figww} shows the non-linear model of the turbine including the surge tank and the travelling wave effects in the penstock, as presented in \cite{demello1992hydraulic}. The model is based on \eqref{hoverqSimpl}, \eqref{hoverqSimpl2} and \eqref{hoverqsurge} and includes the losses in the penstock, the tunnel and the surge tank. It should be noted that the $e^{-2T_es}$ term in \eqref{tanh} is a time delay in the time domain, ensuring a simple representation of the penstock dynamics.

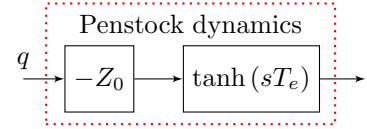
\begin{figure}[!t]
\centering
\begin{tikzpicture}[auto, node distance=1cm,>=latex']
    \node [input, name=qinput] (qinput) {};
    \node [block, right of=qinput, node distance=1cm] (Z0) {$-Z_0$};
    \node [block, right of=Z0, node distance=2cm] (tanh) {$\tanh{(sT_{e})}$};
    \node [output, right of=tanh, node distance=1.5cm] (output) {};
    
    \draw [->] (qinput) -- node[anchor=south,pos=0.0]{$q$} (Z0);
    \draw [->] (Z0) -- (tanh);
    \draw [->] (tanh) -- (output);
    
    \draw[red,thick,dotted]  ($(Z0)+(-0.7,-0.6)$)  rectangle ($(tanh)+(1.1,1)$) ;

    \node [align=left] at ($(Z0)+(1.2,0.7)$) {Penstock dynamics};

;
\end{tikzpicture}
\caption{Penstock model for small signal analysis.} \label{figpd}
\end{figure}

For small signal stability analysis, the penstock dynamic is modelled as shown in Figure \ref{figpd}, using a 4th order approximation of $\tanh$ as given in \eqref{tanh}.

\begin{figure*}[!htb]
\centering
\begin{tikzpicture}[auto, node distance=1cm,>=latex']

    \node [block] (x2) {$x^2$};
    \node [sum,right of =x2, node distance =1cm] (sum1){};
    \draw [->] (x2) --  (sum1);
    \node [block,below of =sum1, node distance =1cm] (cons){$1$};
    \draw [->] (cons) -- node[pos=0.7] {$-$} (sum1);
    \node [block,right of =sum1, node distance =1cm] (sigma){$\sigma$};
    \draw [->] (sum1) --  (sigma);
    \node [sum,right of =sigma, node distance =1cm] (sum2){};
    \draw [->] (sigma) -- node[pos=0.7]{$-$} (sum2);
    \node [block,right of =sum2, node distance =1.6cm] (int){$\frac{1}{s T_{w}}$};
    \draw [->] (sum2) --  (int);
    \node [block,right of =int, node distance =2.8cm] (Q){$\frac{Q_{RT}}{Q_R}$};
    \draw [->] (int) --  node[anchor=south,pos=0.5]{$q_t$} (Q);
    \node [output, right of=Q, node distance=1cm] (outq) {};

    \node [point, below of =int,node distance=1cm] (intb){};
    \node [block,right of =intb, node distance =0.8cm] (pd){$\times/\div$};
    \node [block,left of =pd, node distance =1.5cm] (x22){$x^2$};
    \node [point, right of =int,node distance=1.6cm] (intr){};
    \draw [->] (intr) |-  ([yshift=0.2cm]pd.east);
    \draw [->] (pd) --  (x22);
    \draw [->] (x22) -| node[pos=0.9] {$-$} (sum2);
    
    \node [block,below of =pd, node distance =1.4cm] (p1){$\times$};
    \node [point,left of =pd, node distance =0.7cm] (pdl){};
    \draw [->] (pdl) |-  ([yshift=0.2cm]p1.west);
    \node [block,right of =p1, node distance =1.5cm] (xi){$\xi$};
    \draw [->] (p1) --  (xi);
    \node [sum,right of =xi, node distance =1cm] (sum3){};
    \draw [->] (xi) -- (sum3);
    \node [block,below of =xi, node distance =1.0cm] (psi){$\psi$};
    \draw [->] (psi) -| node[pos=0.9]{$-$}  (sum3);
    
    \node [point, left of =p1,node distance=1.2cm] (p1l){};
    \node [sum,below of =p1l, node distance =0.2cm] (sum4){};
    \draw [->] (sum4) --  ([yshift=-0.2cm]p1.west);
    \node [block,left of =sum4, node distance =1.2cm] (tan){$\tan \alpha_1$};
    \draw [->] (tan) -- (sum4);
    \node [block,left of =tan, node distance =1.5cm] (sin){$\sin x$};
    \draw [->] (sin) -- (tan);
    \node [block,left of =sin, node distance =1.5cm] (alpha){$\alpha_1$};
    \draw [->] (alpha) -- (sin);
    \node [input, left of=alpha, node distance =1.0cm] (inputg) {};
    \draw [->] (inputg) -- node[anchor=north,pos=0.3]{$\kappa$} (alpha);
    \node [input, left of=x2, node distance =1.0cm] (inputw) {};
    \draw [->] (inputw) -- node[anchor=south,pos=0.3]{$\omega$} (x2);
    
    \node [point,right of =inputg, node distance =0.3cm] (inputgr){};
    \node [point,right of =alpha, node distance =0.7cm] (alphar){};
    \node [point,below of =intr, node distance =1.7cm] (intrb){};
    \draw [->] (inputgr) |-  (intrb) |- ([yshift=-0.2cm]pd.east);

    \node [point,right of =inputw, node distance =0.3cm] (inputwr){};
    
    \node [point,right of =psi, node distance =3.5cm] (psir){};
    \node [point,above of =psir, node distance =0.6cm] (ny){};
    \node [block,above of =ny, node distance =0.7cm] (pd2){$\times/\div$};
    \node [point,below of =sin, node distance =1.0cm] (sina){};
    \node [block,right of =sina, node distance =1.0cm] (cos){$\cos{x}$};
    \draw [->] (alphar) |- (cos);
    \draw [->] (cos) -| (sum4);
    
    \node [block,right of =ny, node distance =2.0cm] (p3){$\times$};
    \draw [->] (pd2) -| (p3);
    \node [point,right of =sum3, node distance =1.0cm] (sum3r){};
    \draw [->] (sum3) -- (sum3r) |- (p3);
    
    \node [block,right of=Q, node distance=2.2cm] (ww) {\begin{tabular}{c} Waterway \\ model\\ \end{tabular}};
    \draw [->] (Q) -- node[anchor=south,pos=0.5]{$q$}(ww);
    \node [block,right of =ww, node distance =2.4cm] (H){$\frac{H_{R}}{H_{RT}}$};
    \draw [->] (ww) --  node[anchor=south,pos=0.5]{$h$} (H);
    \node [point,right of =H, node distance =1.5cm] (Hr){};
    \node [point,above of =Hr, node distance =1.0cm] (Hra){};
    \draw [->] (H)  -- node[anchor=south,pos=0.5]{$h_t$}(Hr) -- (Hra) -|  (sum2);
    
    \node [input, left of=psi, node distance =1.0cm] (inputw2) {};
    \draw [->] (inputw2) -- node[anchor=south,pos=0.3]{$\omega$} (psi);
    
    \node [output, right of=p3, node distance=1cm] (outp) {};
    \draw [->] (p3) -- node[anchor=south,pos=0.7]{$P_m$} (outp);
    
    \node [point,right of =pd, node distance =0.8cm] (pdr){};
    \node [point,above of =pdr, node distance =0.2cm] (pdra){};
    \node [point,right of =pdra, node distance =2.5cm] (pdrar){};
    \node [point,below of =pdrar, node distance =1.1cm] (pdrarb){};
    
    \draw [->] (pdra) -- (pdrar) |- ([yshift=0.2cm]pd2.west);

    \node [point,below of =Hr, node distance =0.8cm] (Hrb){};
    \node [point,left of =Hrb, node distance =4.0cm] (Hrbl){};
    \draw [->] (Hr) -- (Hrb) -- (Hrbl) |- ([yshift=-0.2cm]pd2.west) ;
    
    \node [point,below of =p3, node distance =1.0cm] (p3b){};
    \node [input, left of=p3b, node distance=1.0cm] (inputw2) {};
    \draw [->] (inputw2) -| node[anchor=south,pos=0.1]{$\omega$} (p3);
    
    \draw[blue,thick,dotted]  ($(inputw)+(-0.1,-1.5)$)  rectangle ($(Q)+(-0.8,0.7)$) ;
    \draw[green,thick,dotted] ($(Q)+(-0.7,-0.7)$)  rectangle ($(H)+(1.3,0.7)$);
    \draw[red,thick,dotted] ($(inputg)+(-0.1,-1.5)$)  rectangle ($(p3)+(1.1,1.2)$);

    \node [align=left] at ($(inputw)+(1.5,0.9)$) {Momentum equation};
    \node [align=center] at  ($(inputw)+(1.2,-3.9)$) {Torque equation};

\end{tikzpicture}
\caption{Turbine model based on the Euler Equations.} \label{figNielsen}
\end{figure*}
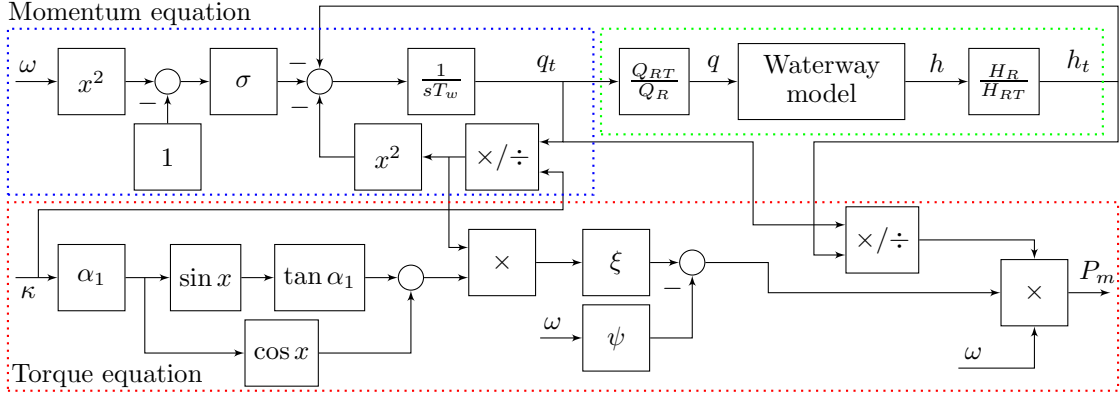

\subsection{Hydraulic Machine Simulation Models}

Hydraulic turbine models have earlier been compared in, for instance, \cite{koritarov2013review}. However, those models are relatively simple and do not consider variations in the turbine rotational speed. This paper presents a hydraulic turbine model based on the Euler turbine equations for grid integration studies of VSHP, and shows how three common hydraulic machine models are simplifications of the Euler equations.


The main part of the Euler turbine equations model is the Euler turbine equations \eqref{turbeuler1} to \eqref{turbeuler3} describing how the hydraulic power is transformed into mechanical rotational power. The model considers both guide vane opening, pressure, water flow and rotational speed and is used together with the waterway model (Figure \ref{figww}) to also consider water hammer, mass oscillation and losses in the waterway, as shown in Figure \ref{figNielsen}. However, it does not consider the effect of acceleration on the flow through the runner and the angular acceleration of the water masses in the runner \cite{nielsen2015simulation}. The dimensionless turbine equations are derived in \cite{nielsen2015simulation} where the dimensionless flow, head and angular speed of rotation are $q_t = Q / Q_{Rt}$, $h_t = H / H_{Rt}$ and $\omega = \Omega / \omega_R$. Since the turbine parameters can be derived from its physical dimensions, no detailed and restricted turbine data are needed. The momentum and torque equations are given as 

\begin{equation}\label{turbeuler1}
    \begin{split}
        T_{w} \frac{dq_t}{dt} &= h_t - \left( \frac{q_t}{\kappa} \right)^2 - \sigma \left( \omega^2 - 1 \right)\\
        T_{a} \frac{d\omega}{dt} &= \frac{1}{h_t} q_t \left(m_s - \psi \omega \right) - P_g
    \end{split}
\end{equation}

\begin{equation}\label{turbeuler2}
    \begin{split}
        m_s &= \xi \frac{q_t}{\kappa} \left( \cos{\alpha_1} + \tan{\alpha_{1R}} \sin{\alpha_1} \right)  \\
        \kappa &= \frac{Q_R}{Q_{Rt}} g , \quad \alpha_1 = \sin^{-1}{ \left(  \kappa \sin{\alpha_{1R} } \right)} 
    \end{split}
\end{equation}

where $P_g$ is the generator power, $\kappa$ the opening degree of the turbine and the turbine parameters $\sigma$, $\psi$, $\xi$ and $\alpha_{1R}$ is found in Appendix \ref{app}. The turbine head $h_t$ and the hydraulic efficiency $\eta_h$ are given as

\begin{equation}\label{turbeuler3}
    \begin{split}
        h_t &= \left( \frac{q_t}{\kappa} \right)^2 - \sigma \left( \omega^2 - 1 \right), \quad    \eta_h = \frac{1}{h} \left( m_s -\psi \omega \right) \omega\\
    \end{split}
\end{equation}

The state-space representation of the \textit{Euler} model is derived by utilising a lumped-parameter equivalent \eqref{tanh} of the penstock dynamics (Figure \ref{figpd}).

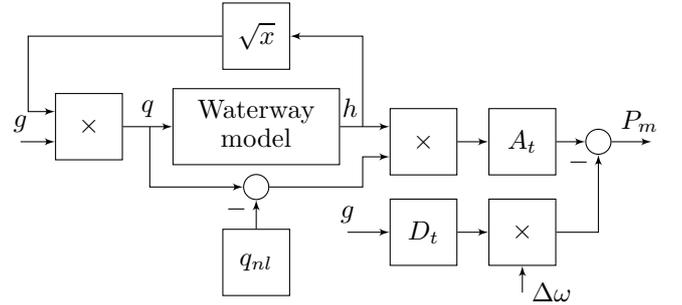
\begin{figure}[!t]
\centering
\begin{tikzpicture}[auto, node distance=1cm,>=latex']
    
    \node [block] (div) {$\times$};
    \node [block,right of=div, node distance=2.2cm] (ww) {\begin{tabular}{c} Waterway \\ model\\ \end{tabular}};
    \draw [->] (div) --  node[anchor=south,pos=0.5]{$q$}(ww);
    
    \node [point, right of =ww,node distance =2.2cm](wwr) {};
    \node [block, below of=wwr,node distance=0.2cm] (p2) {$\times$};
    \draw [->] (ww) -- node[anchor=south,pos=0.2]{$h$}  ([yshift=0.2cm]p2.west);
    
    \node [sum, below of=ww, node distance=0.8cm] (sum2) {};
    \node [block,below of=sum2, node distance=1cm] (k2) {$q_{nl}$};
    \draw [->] (k2) --  node[pos=0.7] {$-$} (sum2);
    \node [point, right of =sum2,node distance =1.4cm](sum2r) {};
    \draw [->] (sum2) -- (sum2r) |- ([yshift=-0.2cm]p2.west);
    \node [point, right of =div,node distance =0.8cm](divr) {};
    \draw [->] (divr) |-  (sum2);
    
    \node [block,right of=p2, node distance=1.3cm] (At) {$A_t$};
    \draw [->] (p2) --  (At);
    \node [sum, right of=At, node distance=1.0cm] (sum3) {};
    \draw [->] (At) --  (sum3);
    \node [output, right of=sum3, node distance=0.7cm] (out) {};
    \draw [->] (sum3) -- node[anchor=south,pos=0.7]{$P_m$} (out);

    \node [block,above of=ww, node distance=1.2cm] (sq) {$\sqrt{x}$};
    \node [point, above of =sum2r,node distance =0.8cm](sum2ra) {};
    \draw [->] (sum2ra) |-  (sq);
    \node [point, left of =div,node distance =0.8cm](divl) {};
    \node [point, above of =divl,node distance =0.2cm](divla) {};
    \draw [->] (sq) -|  (divla) -- ([yshift=0.2cm]div.west);
    \node [point, left of =div,node distance =0.9cm](divl2) {};
    \node [input, below of =divl2 ,node distance =0.2cm] (g) {};
    \draw [->] (g) -- node[anchor=south,pos=0]{$g$} ([yshift=-0.2cm]div.west);
    
    \node [block, below of=At, node distance=1.2cm] (ti) {$\times$};
    \draw [->] (ti) -| node[pos=0.95] {$-$} (sum3);
    
    \node [block, left of=ti, node distance=1.3cm] (D) {$D_t$};
    \draw [->] (D) --  (ti);
    
    \node [input, left of=D, node distance=1.0cm] (pla) {};
    \draw [->] (pla) -- node[anchor=south,pos=0]{$g$} (D);
    
    \node [input, below of=ti, node distance=0.8cm] (tib) {};
    \draw [->] (tib) -- node[anchor=west,pos=0]{$\Delta \omega$} (ti);
    
\end{tikzpicture}
\caption{\textit{IEEE} model} \label{figIEEE}
\end{figure}

The \textit{IEEE} model \cite{demello1992hydraulic} shown in Figure \ref{figIEEE}  utilises the same waterway model as the Euler turbine equations model but has a simpler turbine model. The mechanical power $P_m$ and the water flow $q$ are given by 

\begin{equation}\label{turbIEEE}
    \begin{split}
        P_m &= A_t h \left(q-q_{nl} \right) - D_t g \Delta \omega\\
        q &= g \sqrt{h} 
    \end{split}
\end{equation}

In an ideal turbine, the mechanical power is proportional to the flow $q$ times the head $h$. In this case, the no-load flow $q_{nl}$ and the factor $A_t$ are added to include a simple loss model. The equation includes the speed damping effect, depending on the guide vane opening $g$. The relationship between the flow $q$ and the guide vane opening $g$ is derived from \eqref{turbeuler1} and \eqref{turbeuler2} by assuming stationary conditions, $\omega=1$, $Q_{Rt}=Q_R$ and thereby $q_t=q$.



\begin{figure}[!t]
\centering
\begin{tikzpicture}[auto, node distance=1cm,>=latex']
    \node [block] (k1) {$1$};
    \node [sum, right of=k1, node distance=1.1cm] (sum1) {};
    \draw [->] (k1) --  (sum1);
    \node [block,right of=sum1, node distance=1.2cm] (int) {$\frac{1}{T_w s}$};
    \draw [->] (sum1) --  (int);
    \node [block,right of=int, node distance=1.8cm] (div) {$\times / \div$};
    \draw [->] (int) -- node[anchor=south,pos=0.5]{$q$} (div);
    \node [block,right of=div, node distance=1.5cm] (x2) {$x^2$};
    \draw [->] (div) --  (x2);
    \node [point, right of =x2,node distance =1.2cm](x2r) {};
    \node [point, above of =sum1,node distance =0.8cm](sum1a) {};
    \draw [->] (x2) -- node[anchor=south,pos=0.5]{$h$} (x2r) |- (sum1a) -- node[pos=0.7] {$-$} (sum1);
    \node [block, below of=x2, node distance=1.0cm] (D) {$D_t$};
    \node [point, below of =div,node distance =1.0cm](divb) {};
    \node [input, left of=divb, node distance=0.4cm] (divbl) {};
    \draw [->] (divb) --  (div);
    \draw [->] (divbl) -- node[anchor=south,pos=0]{$g$} (D);
    \node [point, right of =int,node distance =0.7cm](intr) {};
    \node [sum, below of=intr, node distance=2.0cm] (sum2) {};
    \draw [->] (intr) --  (sum2);
    \node [block,below of=sum2, node distance=1.0cm] (p) {$\times$};
    \draw [->] (sum2) --  (p);
    \node [block,right of=p, node distance=1.8cm] (At) {$A_t$};
    \draw [->] (p) --  (At);
    \node [sum, right of=At, node distance=1.8cm] (sum3) {};
    \draw [->] (At) --  (sum3);
    \node [block,above of=sum3, node distance=1.0cm] (p2) {$\times$};
    \draw [->] (p2) -- node[pos=0.5] {$-$} (sum3);
    \draw [->] (D) -|  (p2);
    \node [output, right of=sum3, node distance=1.0cm] (out) {};
    \draw [->] (sum3) -- node[anchor=south,pos=0.7]{$P_m$} (out);
    \node [input, left of=p2, node distance=1.0cm] (p2l) {};
    \draw [->] (p2l) -- node[anchor=south,pos=0]{$\Delta \omega$} (p2);
    \node [block,left of=sum2, node distance=1cm] (k2) {$q_{nl}$};
    \draw [->] (k2) --  node[pos=0.5] {$-$} (sum2);
    \node [point, left of =sum1a,node distance =1.8cm](sum1al) {};
    \draw [->] (sum1a) -- (sum1al) |- (p);
;
\end{tikzpicture}
\caption{\textit{Hygov} model} \label{figHygov}
\end{figure}
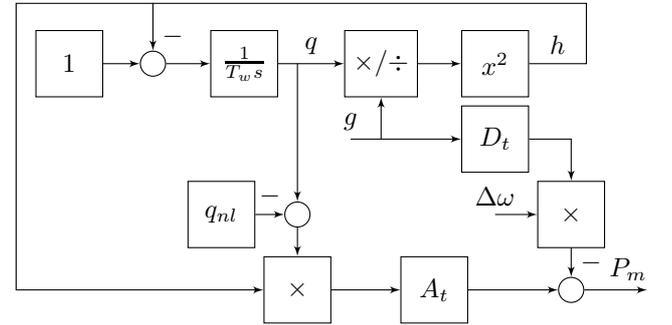

The \textit{Hygov} model presented in \eqref{turb2} and Figure \ref{figHygov} assumes an inelastic water column, does not consider water hammer, mass oscillation or deviation in the rotational speed, and has a simplified relationship between the guide vane opening and the torque. \cite{kundur1994power}. The flow $q$ is found by integration of the turbine head. This equation can be derived from \eqref{turbeuler1} by assuming $\omega = 1$, setting $h_t = 1$ and utilising that $q = g \sqrt{h}$ by assuming no turbine losses. 

\begin{equation}\label{turb2}
    \begin{split}
        P_m &= A_t h \left(q-q_{nl} \right) - D_t g \Delta \omega\\
        \frac{dq}{dt} &= \frac{1}{T_w}\left(1-h \right), \quad h = \left( q/g \right)^2
    \end{split}
\end{equation}

\begin{figure}[!t]
\centering
\begin{tikzpicture}[auto, node distance=1cm,>=latex']
    \node [input] (ginput) {};
    \node [block, right of=ginput, node distance=1.5cm] (Z) {$\frac{1-T_w s}{1 + \frac{1}{2}T_w s}$};
    \node [output, right of=Z, node distance=1.5cm] (output) {};
    
    \draw [->] (ginput) -- node[anchor=south,pos=0.0]{$g$} (Z);
    \draw [->] (Z) -- node[anchor=south,pos=0.7]{$P_m$}(output);
\end{tikzpicture}
\caption{\textit{Linearised} hydraulic turbine model} \label{figSimpleTurbine}
\end{figure}
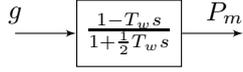

The linearised hydraulic turbine model presented in \eqref{turb1} and Figure \ref{figSimpleTurbine} only considers the water starting time $T_w$ \cite{kundur1994power}. The model is a linearisation of the \textit{Hygov} model around the nominal operating point but without the speed damping term. 

\begin{equation}\label{turb1}
    \begin{split}
        \frac{P_m (s)}{g(s)} = \frac{1-T_w s}{1 + \frac{1}{2}T_w s}
    \end{split}
\end{equation}

\subsection{Control Objectives and Design}

The suggested main control objectives for the VSHP are:

\begin{itemize}
    \item Control objectives for internal power plant control:
    \begin{itemize}
        \item Optimise the rotational speed of the turbine with respect to the efficiency at part load
        \item Minimise water hammering and mass oscillations
        \item Minimise guide vane servo operation
        \item Minimise hydraulic and electric losses
    \end{itemize}
    \item Control objectives for the provision of power system support (ancillary services):
    \begin{itemize}
        \item Contribute to primary frequency regulation
        \item Increase the system inertia by virtual inertia control
        \item Increase the voltage control speed
    \end{itemize}
\end{itemize}

The achievement of the internal control objectives will to a great extent decide the constraints in reaching the power system support objectives. This paper will, therefore, primarily concentrate on modelling of the VSHP for achieving the internal control objectives. 

\subsection{Governor PID Controller without Permanent Droop}

\begin{figure}[!t]
\centering
\begin{tikzpicture}[auto, node distance=1cm,>=latex']
    
    \node [sum] (sum) {};
    \node [input, left of=sum, node distance=0.8cm] (in) {};
    \draw [->] (in) -- node[anchor=south,pos=0.2]{$\omega^*$}  (sum) ;
    \node [input, below of=sum, node distance=0.6cm] (in2) {};
    \draw [->] (in2) -- node[anchor=north,pos=0]{$\omega$} node[pos=0.8] {$-$} (sum) ;
    
    \node [block, right of=sum, node distance=1.7cm] (d) {PID};
    \draw [->] (sum) -- node[pos=0.5]{$\Delta \omega$}(d);
    
    \node [block, right of=d, node distance=1.9cm] (g) {$\frac{1}{1+T_G s}$};
    \draw [->] (d) -- node[pos=0.5]{$g^*$}(g);
    
    \node [output, right of=g, node distance=1.5cm] (outd) {};
    \draw [->] (g) -- node[anchor=south,pos=0.7]{$g$} (outd);
    
    \node [point, above of=d, node distance=0.7cm] (da) {};
    \node [point, right of=da, node distance=0.2cm] (dar) {};
    \node [point, right of=dar, node distance=0.2cm] (darr) {};
    \draw [-] (d) --  (dar);
    \draw [-] (dar) --  (darr);
    
    \node [point, below of=d, node distance=0.7cm] (db) {};
    \node [point, left of=db, node distance=0.2cm] (dbl) {};
    \node [point, left of=dbl, node distance=0.2cm] (dbll) {};
    \draw [-] (d) --  (dbl);
    \draw [-] (dbl) --  (dbll);
    
\end{tikzpicture}
\caption{Governor with PID control and without droop} \label{figgov}
\end{figure}
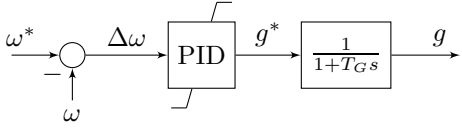

A governor with a PID controller and without permanent droop is presented in Figure \ref{figgov} and \eqref{eq:gov}. Droop control is not required since the grid converter performs power control. The PID controller output is saturated and rate limited.

\begin{equation}\label{eq:gov}
    \begin{split}
        \frac{g}{\Delta \omega} = \frac{g}{\omega^*-\omega} = \frac{k_{g,d} s^2+k_{g,p} s+k_{g,i}}{s} \frac{1}{1+T_G s}
    \end{split}
\end{equation}

\section{Comparison of Turbine Models}\label{ComparisonModels}

\begin{figure}[!t]
\centering
    \includegraphics[scale=0.8]{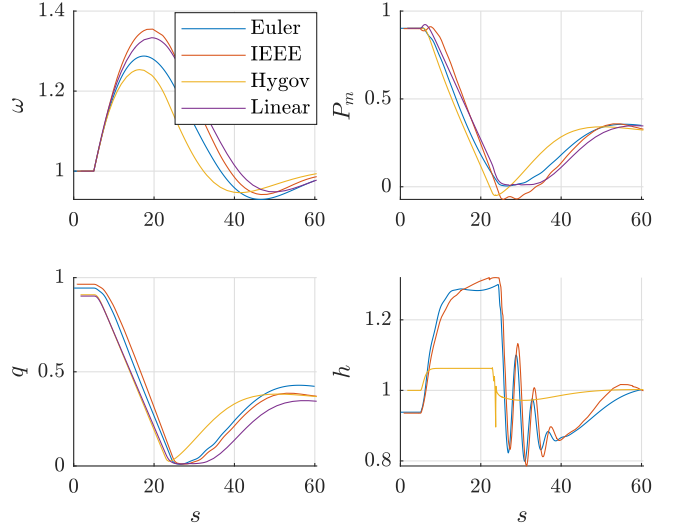}
    \caption{Comparison of turbine dynamics - step on $P^*$ from 0.9 to 0.3~p.u. at $t=5s$.} \label{figStepP}
\end{figure}

\begin{figure}[!t]
\centering
    \includegraphics[scale=0.8]{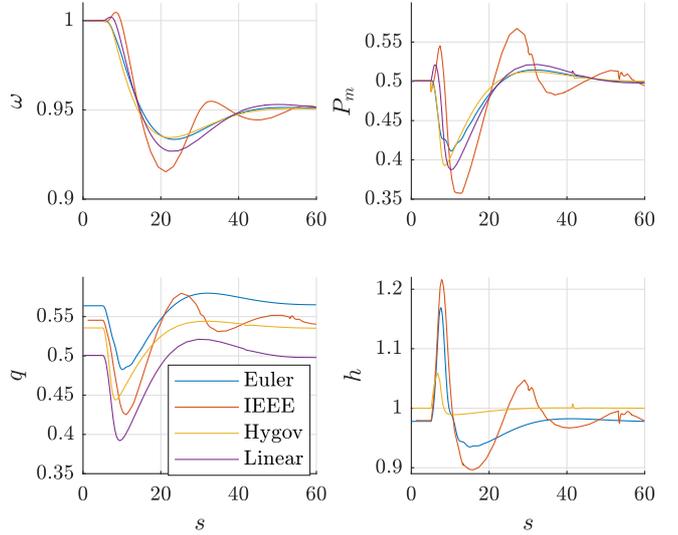}
    \caption{Comparison of turbine dynamics - step on $\omega^*$ from 1.00 to 0.95~p.u. at $t=5s$.} \label{figStepOmega}
\end{figure}

\begin{figure}[!t]
\centering
    \includegraphics[scale=0.8]{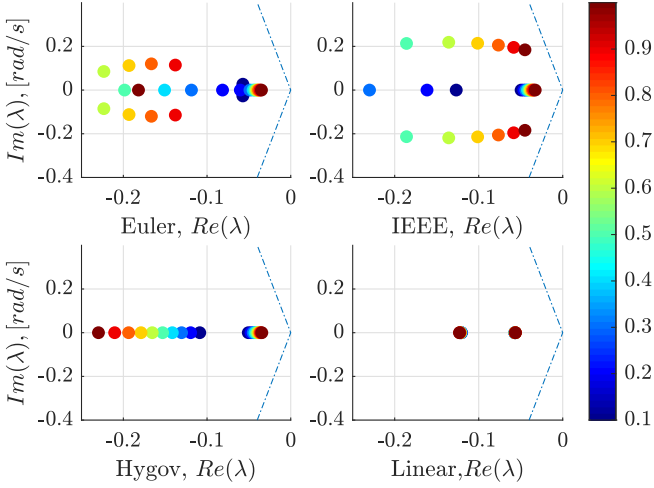}
    \caption{Comparison of eigenvalues at different power references $P^*$ and $\omega^*=1.0$, where $P^*$ is represented by the colour scheme.} \label{figEigP}
\end{figure}

\begin{figure}[!t]
\centering
    \includegraphics[scale=0.8]{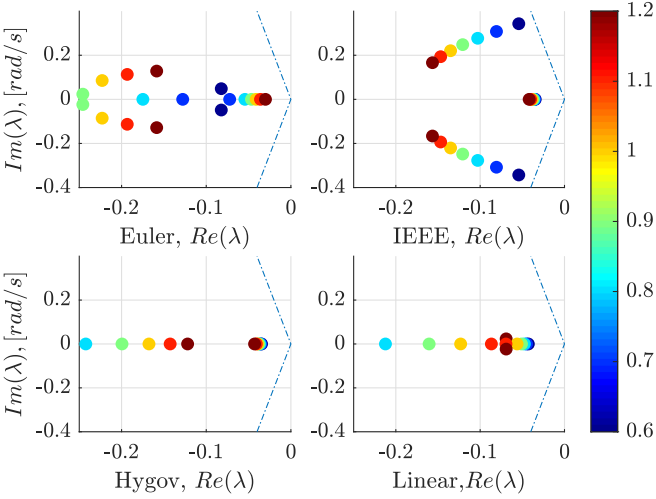}
    \caption{Comparison of eigenvalues at different turbine rotational speed references $\omega^*$ and $P^*=0.6$, where $\omega^*$ is represented by the colour scheme.} \label{figEigOmega}
\end{figure}

\begin{figure}[!t]
\centering
    \includegraphics[scale=0.8]{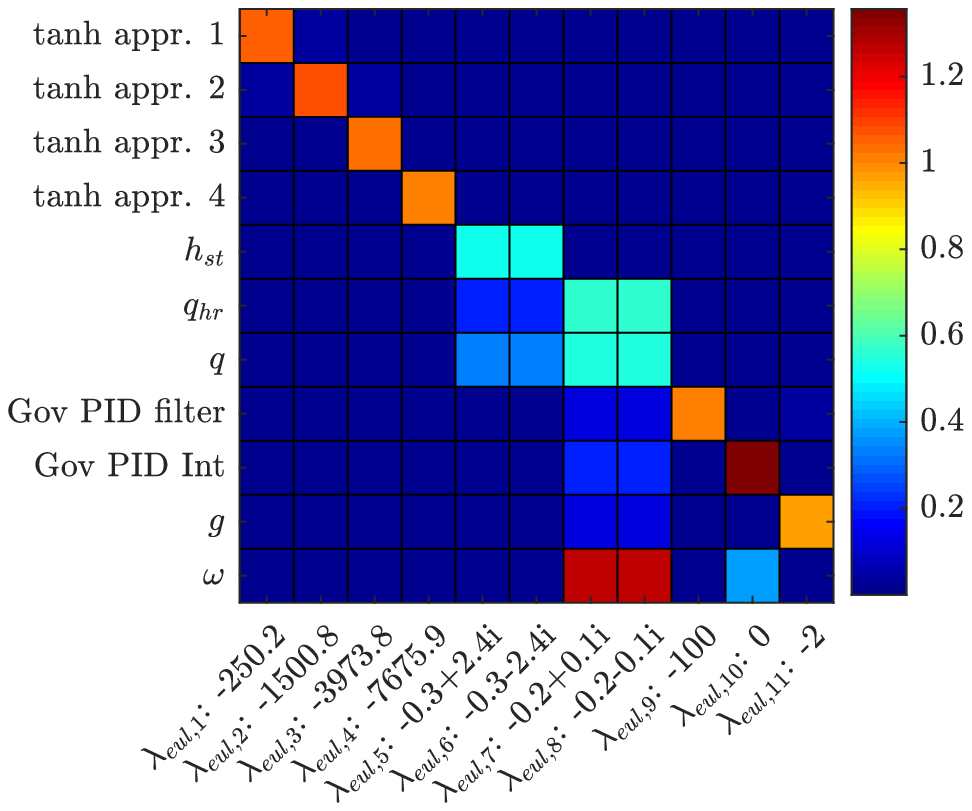}
    \caption{Participation factor matrix for the \textit{Euler} model with system inputs $\omega^* = 1.0$ and $P^*=0.6$. The colour represents the absolute value of the relative participation of the state variables in the modes.}\label{pf_nie}
\end{figure}

\begin{figure}[!t]
\centering
    \includegraphics[scale=0.8]{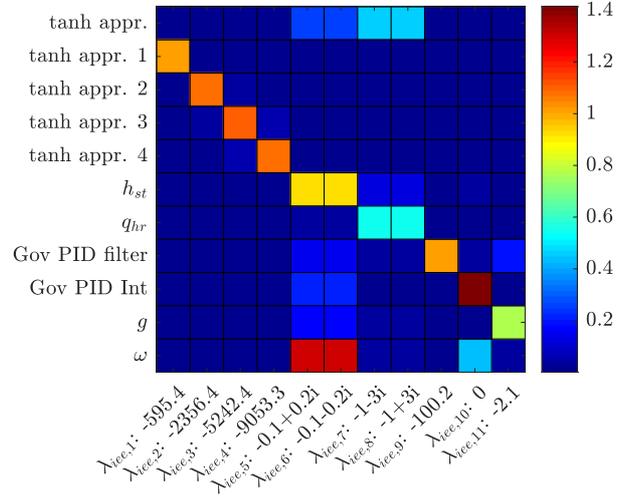}
    \caption{Participation factor matrix for the \textit{IEEE} model with system inputs $\omega^* = 1.0$ and $P^*=0.6$. The colour represents the absolute value of the relative participation of the state variables in the modes.}\label{pf_iee}
\end{figure}

The demand for modelling of the turbine is different for a VSHP compared to a conventional hydropower plant because of the large variance and dynamics in turbine rotational speed. Dynamic and eigenvalue analyses are performed on four different turbine models with different levels of detail to find a suitable turbine model for grid integration studies of VSHP. Figures \ref{figStepP} and \ref{figStepOmega} show the step response of the VSHP after a step in, respectively, reference power $P^*$ and reference turbine rotational speed $\omega^*$ for the four different turbine models. The corresponding eigenvalue plots for different values of $P^*$ and $\omega^*$ in the particular operational points are shown in Figures \ref{figEigP} and \ref{figEigOmega}. Together with the participation matrices in Figures \ref{pf_nie} and \ref{pf_iee}, they are the basis for explaining how the models handle a varying turbine rotational speed. A selection of the most important modes related to the turbine models is provided in the participation factor matrices and the colour represents the absolute value of the relative participation of the state variables in the modes.

\begin{figure}[!t]
\centering
    \includegraphics[scale=0.8]{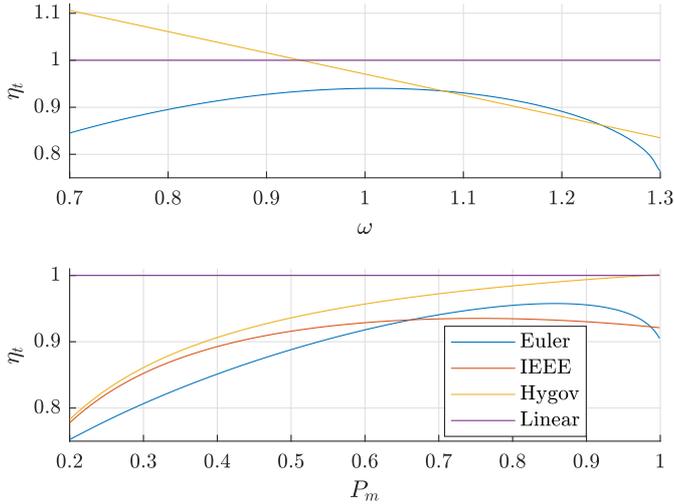}
    \caption{Hydraulic efficiency as function of turbine rotational speed $\omega$ and turbine power $P_m$. The efficiencies of the \textit{IEEE} and \textit{Hygov} models as function of $\omega$ are equal.}\label{fig:eff}
\end{figure}

The \textit{Euler} and \textit{IEEE} models have a mode pair and a single-pole related to the governor control loop; $\lambda_{eul,7-8}$ and $\lambda_{eul,10}$ for the \textit{Euler} model and $\lambda_{iee,5-6}$ and $\lambda_{iee,10}$ for the \textit{IEEE} model. The frequency of the governor mode pair is approximately 0.02~Hz and is recognised in Figures \ref{figStepP} and \ref{figStepOmega} as the oscillations with the largest amplitude and as the only mode pair in Figures \ref{figEigP} and \ref{figEigOmega}. For the \textit{Euler} and \textit{IEEE} models, the eigenvalue plots in Figure \ref{figEigP} show that the relative damping of these modes decreases with increasing power output $P^*$. This can be explained by the hydraulic efficiency, as shown in Figure \ref{fig:eff}; the relative losses are higher for cases with low power, and thereby the damping is better for these cases. Figure \ref{figEigOmega} shows that the damping of the \textit{IEEE} modes related to governor control is less at low rotational speed $\omega$. This connection is more complex for the \textit{Euler} modes related to governor control; the damping is less for high and low turbine rotational speed and higher for nominal rotational speed. The governor parameters must, therefore, be tuned for the worst-case operating points; the combinations of the minimum and maximum power and rotational speed.

The \textit{Linearised} and \textit{Hygov} models do not have any mode pair related to the governor control loop for most operating points. Generally, the lowest damping for these models is found for low power and high rotational speed.



The \textit{Euler} and \textit{IEEE} models include dynamics of the penstock, the headrace tunnel and the surge tank. The modes $\lambda_{eul,5-6}$ and $\lambda_{iee,7-8}$ primarily relate to the surge tank head $h_{st}$, the headrace tunnel flow $q_{hr}$ and the penstock flow $q$ ($\tanh$-approximation for the \textit{IEEE} model), as seen from Figures \ref{pf_nie} and \ref{pf_iee}. These modes correspond to oscillations between the turbine and the surge tank and can be recognised as oscillations at approximately 0.4~Hz in the turbine head $h$ in Figure \ref{figStepP}. 

The \textit{Euler} and \textit{IEEE} models allow elastic waterway and water hammering. The modes $\lambda_{eul,1-4}$ and $\lambda_{iee,1-4}$ are related to this phenomena, all of them related to the states in the lumped-parameter equivalent for $\tanh{sT_e}$ \eqref{tanh}, as observed from Figures \ref{pf_nie} and \ref{pf_iee}. The water hammering is well damped when the guide vane closing time is longer than twice the elastic water time constant for the penstock. This is the case in Figures \ref{figStepP} and \ref{figStepOmega} where no water hammering is observed. If the guide vane opening time is faster than the reflection time of the pressure wave in the penstock, water hammering may occur.

The total hydraulic efficiency is presented as functions of, respectively, the turbine rotational speed $\omega$ and the turbine power $P_m$ in Figure \ref{fig:eff}. The efficiency of the \textit{IEEE} model will, in contrast to the \textit{Hygov} model, decrease for high turbine power since the waterway losses are included. As the rotational speed increases, the damping term of these two models will result in a decreasing efficiency. The efficiency curve of the \textit{Euler} model is closer to the reality since it has a maximum efficiency point near the nominal rotational speed, mainly due to the $\sigma$-parameter. However, the parameters of the \textit{Euler} model must be tuned to represent the hill diagram of the turbine in the best way. 

The most detailed model is the \textit{Euler} model, considering many aspects of the turbine:
\begin{itemize}
    \item The flow $q$ is dependent on the rotational speed $\omega$ 
    \item The torque and the efficiency are dependent on the rotational speed $\omega$ and they are nonlinear functions of the flow $g$.
\end{itemize}

Another advantage of the \textit{Euler} model is that its parameters can be found directly from the turbine dimensions.

The torque calculated by the \textit{Euler} equations will increase with decreasing rotational speed $\omega$. However, as $P_m = T_m \omega$, the output power will reduce. This may cause problems in cases where the output power to the grid is high and the rotational speed is decreasing below a critical value. In this situation, the turbine will not be able to deliver enough power to regain the nominal rotational speed and the turbine will stop if the output power to the grid is not reduced quickly. 


\section{Conclusion}\label{Conclusion}

The analysis and design of hydropower conversion and control systems require accurate modelling of the turbine mechanical power and rotational speed. Precise models are needed when developing a control design aiming to maximise the utilisation of the rotational energy in the generator and the turbine for the provision of fast frequency reserves during power system disturbances. 
In this paper, four hydraulic turbine models are investigated to show that a detailed model is needed for grid integration studies of VSHP. A hydraulic model based on the Euler turbine equations including a one-dimensional waterway model is proposed for the purpose.

When linearised, all four models have a similar complex mode pair and a real pole related to the governor control loop. The relative damping of the complex mode pair reduces with higher power levels and lower rotational speeds. The \textit{Euler} and the \textit{IEEE} models add dynamics related to the penstock, the headrace tunnel and the surge tank, and mass oscillations are recognised in the head and mechanical power. 

The \textit{Linearised} model is independent of both the power level and the turbine rotational speed. The \textit{Hygov} model has a very simplified model of the waterway and the head, and thereby the mechanical power during transients becomes inaccurate. A more detailed one-dimensional waterway model is included in the \textit{IEEE} model; however, the simple turbine model does not consider that the turbine efficiency depends on the rotational speed and is a non-linear function of the flow. For the \textit{Euler} model, the Euler turbine equations are used directly to describe how the hydraulic power is transformed into mechanical power. This includes the dependency between the water flow and the rotational speed, and the representation of losses is more detailed. This is also the only model that considers that the turbine power will be reduced as the rotational speed reduces. The investigation of the turbine models shows that the variable speed operation of the hydropower plant causes the need for a detailed hydraulic model. Therefore, our conclusion is that the accuracy of the \textit{Euler} model is needed for simulating the transients and variation in rotational speed that will take place in a VSHP. Further analysis will answer how this model interacts with the rest of the power system.


\bibliography{Paper1}             
                                                   







\appendix
\section{Parameters, Set-points and Variables}\label{app}    


\begin{table}[hb]
\label{table_parameters}
\caption{Parameters and set-points}
\centering
\begin{tabularx}{0.9\linewidth}{|Xll|}
\hline
\textbf{Parameter} & \textbf{Symbol} & \textbf{Value}  \\
\hline
\multicolumn{3}{|l|}{\textbf{Waterway}}\\
Rated water flow & $Q_R$ $[m^3/s]$  & 170\\
Rated height & $H_R$ $[m]$  & 425\\
\multicolumn{3}{|l|}{\textit{Penstock:}}\\
Water starting time & $T_w$ $[s]$  & 1.211\\
Water traveling time & $T_e$ $[s]$  & 0.126\\
Characteristic impedance & $Z_0$ $[pu]$ & 9.61\\
Friction factor & $f_{p1}$ $[pu]$   & 0.049\\
\multicolumn{3}{|l|}{\textit{Surge tank:}}\\
Friction factor & $f_{p0}$ $[pu]$  & 0.036\\
Storage constant & $C_{s}$ $[pu]$ & 0.099\\
\multicolumn{3}{|l|}{\textit{Head race tunnel:}}\\
Water starting time & $T_{w2}$ $[s]$   & 4.34\\
Friction factor & $f_{p2}$ $[pu]$  & 0.020\\
\multicolumn{3}{|l|}{\textbf{Hydraulic Machine}} \\
Turbine gain & $A_t$ $[pu]$ &  1.075 \\
No-load water flow & $q_{nl}$ $[pu]$ &  0.07 \\
Turbine mechanical damping & $D_t$ $[pu]$ & 0.5 \\
Turbine constant & $\psi$ $[pu]$ & 0.376\\
Turbine constant & $\xi$ $[pu]$ & 0.906 \\
Turbine constant & $\alpha_{1R}$ $[pu]$ & 0.738 \\
Turbine constant & $\sigma$ $[pu]$ & 0.369 \\
Rated speed & $\Omega$ $[rpm]$  & 750 \\
Rated water flow & $Q_{Rt}$ $[m^3/s]$ & 153 \\
Rated height & $H_{Rt}$ $[m]$ & 425 \\
\multicolumn{3}{|l|}{\textbf{Governor}} \\
Rotational speed reference & $\omega^*$ $[pu]$  & 1.00\\
Governor proportional gain & $k_{g,p}$ $[pu]$ & 1.80\\
Governor integration gain & $k_{g,i}$ $[pu]$ & 0.172\\
Governor derivation gain & $k_{g,d}$ $[pu]$ & 0.696\\
Rate limit & $[pu/s]$ & +/-0.05\\
Servo time constant & $T_G$ $[s]$   & 0.500 \\
\hline
\end{tabularx}
\end{table}

\begin{table}[hb]
\begin{center}
\caption{Variables}
\label{table_variables}
\centering
\begin{tabularx}{0.67\linewidth}{|Xl|}
\hline
\textbf{Variable} & \textbf{Symbol} \\
\hline
\multicolumn{2}{|l|}{\textbf{Waterway}} \\
Surge tank head &  $h_{st}$ \\
Head race tunnel flow & $q_{hr}$ \\
\multicolumn{2}{|l|}{\textbf{Hydraulic Machine}} \\
Turbine head &  $h$ \\
Turbine water flow & $q$ \\
Mechanical torque & $T_m$ \\
Mechanical power & $P_m$ \\
Turbine efficiency & $\eta_h$ \\
Turbine head & $h_t$\\
Turbine flow & $q_t$\\
Opening degree of turbine & $\kappa$\\
\multicolumn{2}{|l|}{\textbf{Governor}}\\
Guide vane opening reference & $g^*$ \\
Guide vane opening & $g$ \\
\hline
\end{tabularx}
\end{center}
\end{table}
\end{document}